\documentclass[conference]{IEEEtran}
\IEEEoverridecommandlockouts

\usepackage[inline]{enumitem}
\usepackage{cite}
\usepackage{url}
\usepackage{hyperref}
\usepackage{amsmath,amssymb,amsfonts}

\makeatletter
\let\MYcaption\@makecaption
\makeatother

\usepackage[font=footnotesize]{subcaption}

\makeatletter
\let\@makecaption\MYcaption
\makeatother

\usepackage{epstopdf}
\usepackage{algorithmic}
\usepackage{graphicx}
\usepackage{textcomp}
\usepackage[dvipsnames,svgnames]{xcolor}
\usepackage[framemethod=tikz]{mdframed}

\def\BibTeX{{\rm B\kern-.05em{\sc i\kern-.025em b}\kern-.08em
    T\kern-.1667em\lower.7ex\hbox{E}\kern-.125emX}}

\usepackage{balance}

\newcounter{insightcounter}
\renewcommand{\theinsightcounter}{\arabic{insightcounter}}

\newmdenv[
  skipabove=1em,
  skipbelow=1em,
  backgroundcolor=lightgray!20, %
  linecolor=darkgray, %
  linewidth=1pt, %
  roundcorner=2pt, %
  innertopmargin=0.8\baselineskip,
  innerbottommargin=0.8\baselineskip,
  innerleftmargin=0.5em,
  innerrightmargin=0.5em,
]{insightbox}

\begin{document}

\title{An~Analysis~of~HPC~and~Edge Architectures~in~the~Cloud\thanks{Parts of this research are funded by the Bavarian Research Institute for Digital Transformation (bidt), an institute
of the Bavarian Academy of Sciences and Humanities.}}

\author{\IEEEauthorblockN{Steven Santillan}
\IEEEauthorblockA{\textit{Escuela Superior Politécnica del Litoral, ESPOL}\\
Guayaquil, Ecuador \\
steisant@fiec.espol.edu.ec}
\and
\IEEEauthorblockN{Cristina L. Abad}
\IEEEauthorblockA{\textit{Escuela Superior Politécnica del Litoral, ESPOL}\\
Guayaquil, Ecuador \\
cabad@fiec.espol.edu.ec}
}

\maketitle

\begin{abstract}
We analyze a recently published dataset of 396 real-world cloud architectures deployed on AWS, from companies belonging to a wide range of industries.
From this dataset, we identify those architectures that contain HPC or edge components and characterize their designs.
Specifically, we investigate the prevalence and interplay of AWS services within these architectures, examine the types of storage systems employed, assess architectural complexity and the use of machine learning services, discuss the implications of our findings and how representative these results are of HPC and edge architectures in the cloud.
This characterization provides valuable insights into current industry practices and trends in building robust and scalable HPC and edge solutions in the cloud continuum, and can be valuable for those seeking to better understand how these architectures are being built and to guide new research.
\end{abstract}

\begin{IEEEkeywords}
HPC, edge computing, fog computing, cloud computing, architectures, workflows, AWS, cloud continuum.
\end{IEEEkeywords}

\section{Introduction}
\label{sec:introduction}
As more types of workloads are moved into the cloud, it is important for practitioners and researchers to have a deep understanding of how these architectures are being built in real-world scenarios.
While general cloud architecture patterns are well-documented (e.g., see~\cite{aws:well:architected:framework}), specific designs for the HPC and edge domains remain less studied.
This gap in knowledge makes it challenging to identify optimal design patterns and for researchers areas requiring further innovation.

Towards the goal of filling this gap, we curate and analyze a subset of the recently released Cloudscape dataset~\cite{Satija:2025:Cloudscape} of real cloud applications extracted from AWS videos in YouTube.
We categorized each of the 396 cloud architectures as HPC, edge, HPC+edge, or none, based on the presence of specific HPC or edge components and services.
To enhance our analysis, we augmented the original Cloudscape dataset with YouTube video publication dates for each architecture, which allowed us to observe trends in the prevalence of HPC and edge architectures over time. 

We then characterize the architectures in our curated dataset, seeking to answer seven research questions covering areas like service use, architectural complexity, use of storage services and machine learning services, and evolution over time. 
We present the results of our analysis and draw insights from these.

We found that the S3 object store is the most popular storage system, even in HPC architectures for which a high-performance storage system, FSx, exists.
For storage, second in popularity is FSx for HPC and NoSQL databases for edge.
For HPC architectures, HPC-specific services like Batch and FSx are used in combination with computing (EC2, Lambda and EKS).
From the edge-specific services, the most frequently used one is the CloudFront CDN, followed by IoT Core;
other edge-specific services like Lambda@Edge and Greengrass are less popular.
In fact, in-house and third-party edge services and devices can be more popular than the AWS-specific ones.
HPC and Edge architectures have a slightly higher architectural complexity than non-HPC, non-edge architectures in the number of services and workflows present in them.

In this work we make the following contributions:
\begin{enumerate}%
    \item We curate and extend the recently released Cloudscape to compile a dataset of real industry HPC and edge architectures running on AWS,
    \item we provide a detailed characterization of real-world HPC and edge cloud architectures deployed on AWS,
    \item we discuss the implications of our findings, offering valuable insights for practitioners and researchers, and
    \item our data and code has been publicly released for others to easily build upon our work.\footnote{See: \url{https://github.com/disel-espol/hpc-and-edge-cloud-architectures/}}
\end{enumerate}

The rest of this paper is organized as follows.
Section~\ref{sec:dataset} describes our curation and dataset enhancement process.
In Section~\ref{sec:analysis} we present the results of our characterization and analysis, and discuss several insights that stem from this study.
In Section~\ref{sec:threats:to:validity} we discuss the limitations of our work (threats to validity).
Section~\ref{sec:related} contextualizes our work within the related work of the community, explaining how we fill an important gap in knowledge regarding real HPC and edge deployments in the cloud.
Finally, in Section~\ref{sec:conclusions} we conclude.

\section{Dataset}
\label{sec:dataset}
\paragraph*{\hspace{-20pt}Data source}
We analyzed a subset of the recently released Cloudscape dataset~\cite{Satija:2025:Cloudscape}.
This dataset has 396 cloud architectures built on AWS, belonging to 378 real businesses, small and large, from a diverse set of industries.
The data was collected by a team from UW-Madison using a systematic process that involved watching videos published by AWS from March 2019 to December 2023 on the ``This is My Architecture'' YouTube playlist, and manually curating and cataloguing the architectures described in the videos.

\vspace{6pt}\paragraph*{\hspace{-20pt}Curation}
We categorized each architecture in Cloudscape as HPC, edge, HPC+edge, or none, depending on whether they contained HPC or edge components/services or not.

We obtained the list of high performance computing (\textbf{HPC}) services from the AWS website~\cite{awsHPC} but removed EC2 from the list, as EC2 virtual machines are also commonly present in Big Data, web and mobile architectures, which are not the focus of our study.
We manually analyzed the data to make sure that excluding EC2 did not lead to the exclusion of a significant number of architectures and found this to be the case, as many HPC architectures that had EC2 VMs also contained other HPC services that helped our script identify them as such.
To this set, we added three architectures that our service-based filtering had missed.
We identified these by doing a keyword search on the titles and descriptions of the architectures in Cloudscape; we searched for \emph{high-performance}, \emph{hpc} and \emph{parallel} and read the descriptions to make sure that the keyword matches were not present for an unrelated reason.
We found that \emph{parallel} was present in a few non-HPC architectures in the context of handling parallel client requests; we did not add these to our filtered dataset.

Similarly, to obtain architectures that contain one of more \textbf{edge} services, we used AWS's list of edge services~\cite{awsEdge} and found those architectures that contained one or more.
In addition, we added architectures that contained an external edge device or service as identified by different tags in the Cloudscape dataset.\footnote{\scriptsize
UserConsumerCamera,
UserCompanyEdge,
UserCompanyDrone,
UserConsumerEdge,
UserConsumerIOT,
UserConsumerPOS,
UserConsumerFarmer,
UserConsumerAlexaGoogleHome,
UserCompanyElementalLiveDevice,	
UserConsumerTV.}
Finally, we manually added five architectures based on their description or third-party components.

\textbf{HPC+edge} are those that have at least one HPC and one edge service or device (including three architectures that were auto-classified as edge plus a manual HPC tag).
We analyzed the architectures in this group independently of the other two categories; i.e., HPC+edge architectures were not analyzed together with the HPC or edge ones, but just in their own  group.
As Table~\ref{tab:curated:dataset} details the size of each group, the reader can easily calculate how average statistics in the characterization section would change if the HPC+edge architectures were considered together with the edge and HPC groups.

\begin{figure}[t]
  \begin{subfigure}{\linewidth}\centering
    \includegraphics[width=0.7\linewidth]{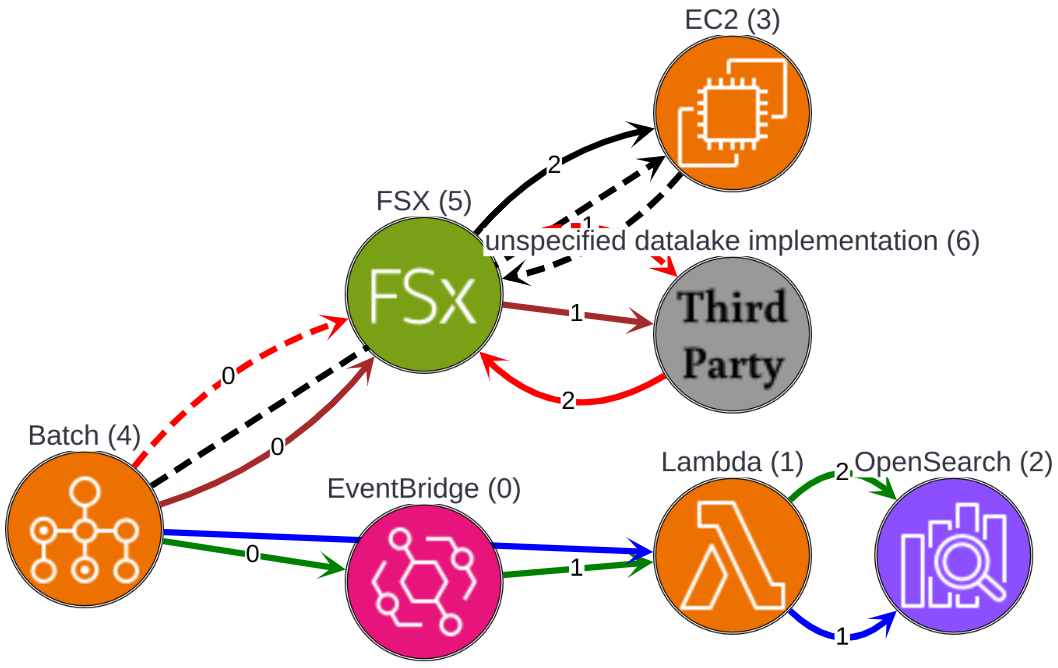}
    \caption{\textbf{HPC.} Neumora Therapeutics:\,Advanced MRI precision medicine and brain data analysis visualization at scale.}
    \label{fig:HPC}
  \end{subfigure}\\
  \begin{subfigure}{\linewidth}\vspace{12pt}
    \includegraphics[width=\linewidth]{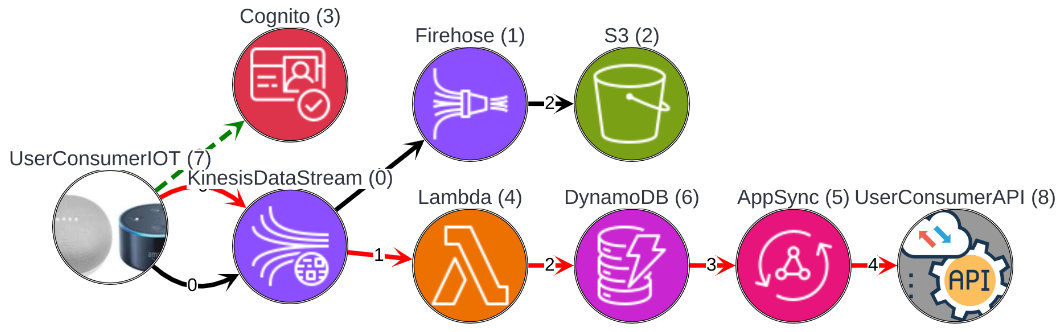}%
    \caption{\textbf{Edge.}\,TechConnect:\,Tech-powered sports with IoT ML.}
    \label{fig:Edge}
  \end{subfigure}\\
  \begin{subfigure}{\linewidth}\vspace{12pt}
    \includegraphics[width=\linewidth]{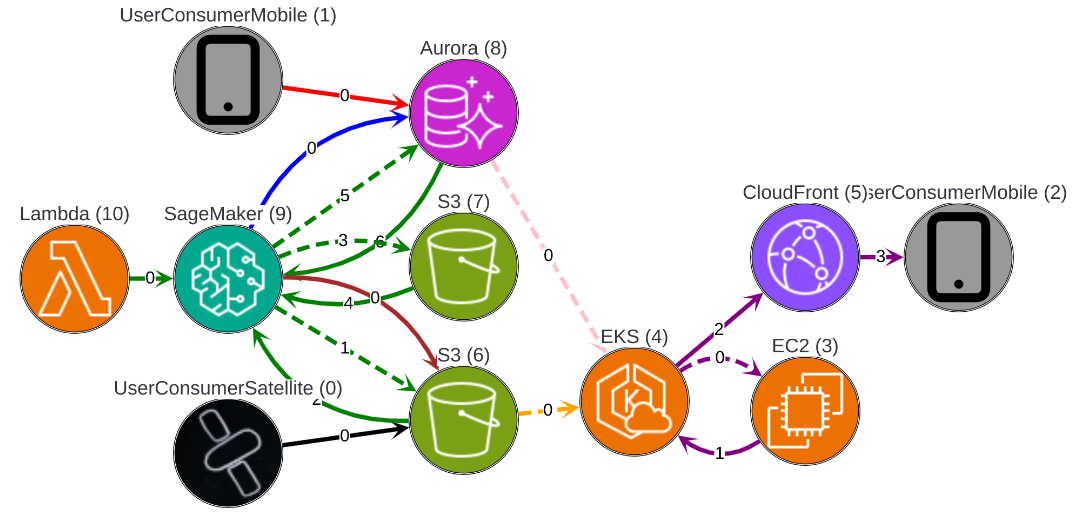}
    \caption{\textbf{HPC+Edge.} TCS:\,GeoSpatial data analysis using TCS digital platform for nextgen agriculture DNA.}
    \label{fig:HPC+Edge}
  \end{subfigure}
  \caption{Three sample architectures. Nodes are services. Edges are interactions between services; solid lines are data interactions, dashed lines are non-data interactions like requests/acks. An architecture may have multiple workflows, each shown with a different color; numbers show the order in the workflow. Nodes with names beginning with \emph{User} represent end user devices or services.}
  \label{fig:sample:architectures}
\end{figure}

For illustrative purposes, Fig.~\ref{fig:sample:architectures} shows one architecture for each of the three categories studied in this paper.\footnote{Images visually rendered with Cloudscape's online Architecture Explorer, available at: \url{https://cloudscape.cs.wisc.edu/}.} 

\begin{figure*}[t]
\centering
  \begin{subfigure}{0.25\textwidth}\centering
    \includegraphics[width=\linewidth]{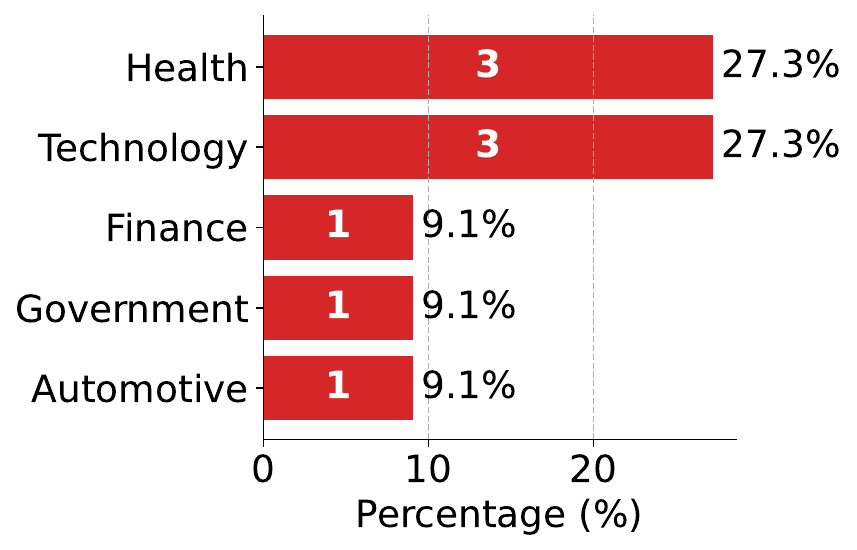}
    \caption{HPC}
  \end{subfigure}%
  \begin{subfigure}{0.25\textwidth}
    \includegraphics[width=\linewidth]{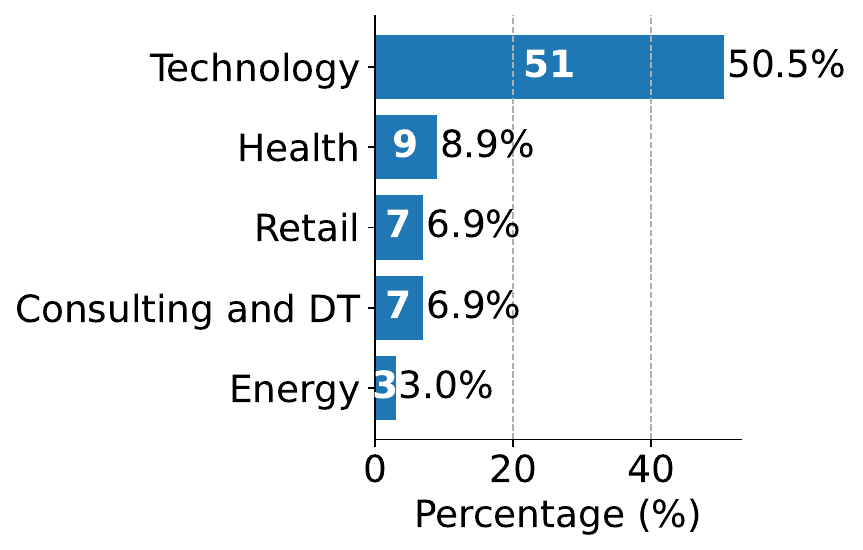}
    \caption{Edge}
  \end{subfigure}%
  \begin{subfigure}{0.25\textwidth}
    \includegraphics[width=\linewidth]{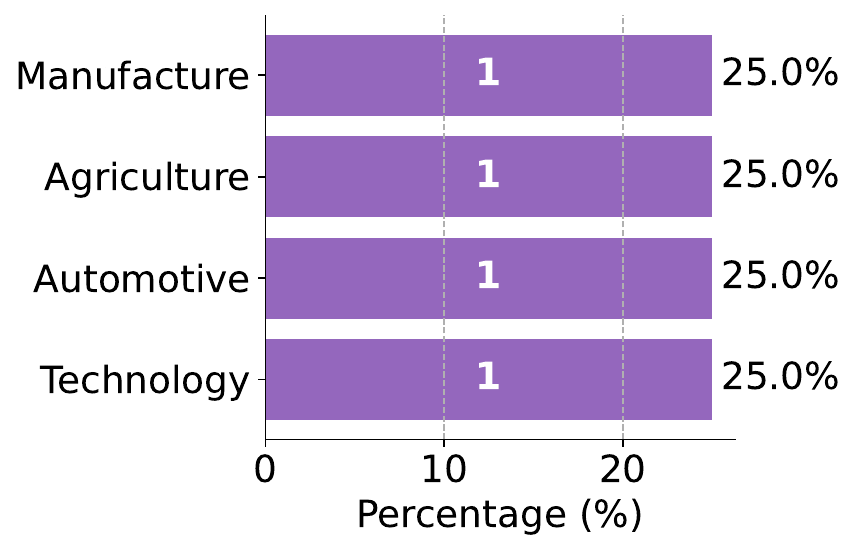}
    \caption{HPC+Edge}
  \end{subfigure}%
  \begin{subfigure}{0.25\textwidth}
    \includegraphics[width=\linewidth]{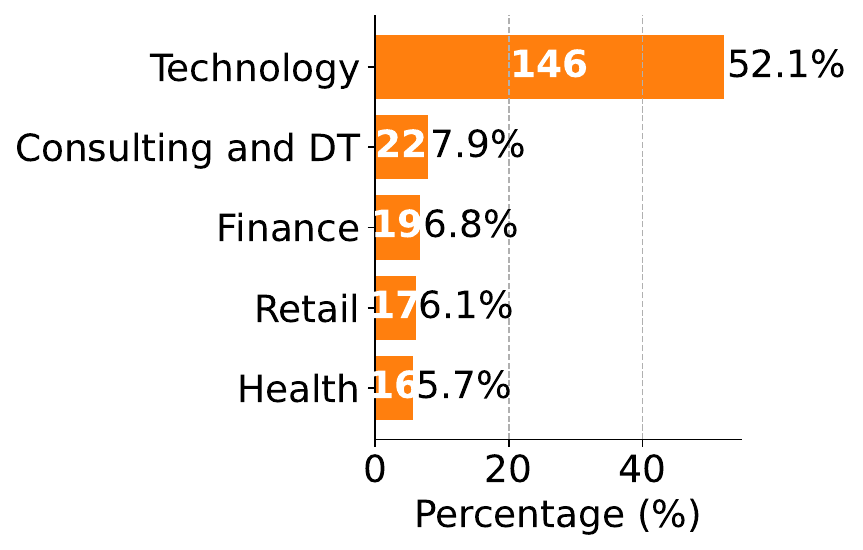}
    \caption{None}
  \end{subfigure}
  \caption{Top five industries to which the architectures in each group belong to. DT: Digital Transformation.}
  \label{fig:industries}
\end{figure*}

\vspace{6pt}\paragraph*{\hspace{-20pt}Augmentation}
The original Cloudscape dataset did not include the YouTube video date.
Our script collects this information for each architecture using YouTube's API and adding a small delay between requests to comply with the API policy.
Adding this information allowed us to observe trends in the published videos per year.
We also manually added an industry category to each architecture (e.g., Tourism, Manufacturing, etc.);
the top 5 industries per type of architecture studied in this paper are shown in Fig.~\ref{fig:industries}.

\vspace{6pt}\paragraph*{\hspace{-20pt}Curated dataset size}
Table~\ref{tab:curated:dataset} shows the number of architectures that fall into each category.
As expected, there are significantly more ($7\times$) edge than HPC architectures.
With only 4 architectures, the HPC+edge group is too small for us to generate interesting insights from it; we report on the results but do not try to make any general claims about them.

\begin{table}[t]
    \centering
    \caption{Prevalence of HPC, edge and HPC+edge architectures in the Cloudscape dataset.}
    \label{tab:curated:dataset}
    \begin{tabular}{l|r|r}
    \hline
    \textbf{Group} & \textbf{Count} & \textbf{Percentage} \\
    \hline
    HPC & 11 & 3\% \\
    Edge & 101 & 26\% \\
    HPC+Edge & 4 & 1\% \\
    None & 280 & 71\% \\
    \hline
    \end{tabular}
\end{table}

\vspace{6pt}\paragraph*{\hspace{-20pt}Data release}
We are publicly releasing our curation, analysis and visualization scripts on \href{https://github.com/disel-espol/hpc-and-edge-cloud-architectures/}{GitHub}.
We believe this aids in reproducibility, allows others to add their own filtering steps, and would help in having an always updated version in case a new version of Cloudscape is released in the future.

\section{Analysis of HPC \& Edge Cloud Architectures}
\label{sec:analysis}
We analyzed our curated dataset, seeking to answer the following research questions:

\newcounter{RQcounter}
\begin{description}
    \item[RQ\refstepcounter{RQcounter}\arabic{RQcounter}:] Which services appear frequently in HPC and edge architectures? Do any co-occur frequently?
    \item[RQ\refstepcounter{RQcounter}\arabic{RQcounter}:] How do the groups differ with respect to the types of services used?
    \item[RQ\refstepcounter{RQcounter}\arabic{RQcounter}:] Which types of storage services appear frequently in HPC and edge architectures?
    \item[RQ\refstepcounter{RQcounter}\arabic{RQcounter}:] Do any of the groups use more machine learning services than the other?
    \item[RQ\refstepcounter{RQcounter}\arabic{RQcounter}:] What are the functional goals of the architectures?
    \item[RQ\refstepcounter{RQcounter}\arabic{RQcounter}:] Is the prevalence of HPC or edge architectures in the original dataset representative of the cloud at large? 
    \item[RQ\refstepcounter{RQcounter}\arabic{RQcounter}:] How many workflows are present in each of the types of architectures studied in the paper?
\end{description}

\subsection{Frequently occurring services (RQ1-RQ2)}
\begin{figure*}[t]
  \begin{subfigure}{0.25\textwidth}\centering
    \includegraphics[width=\linewidth]{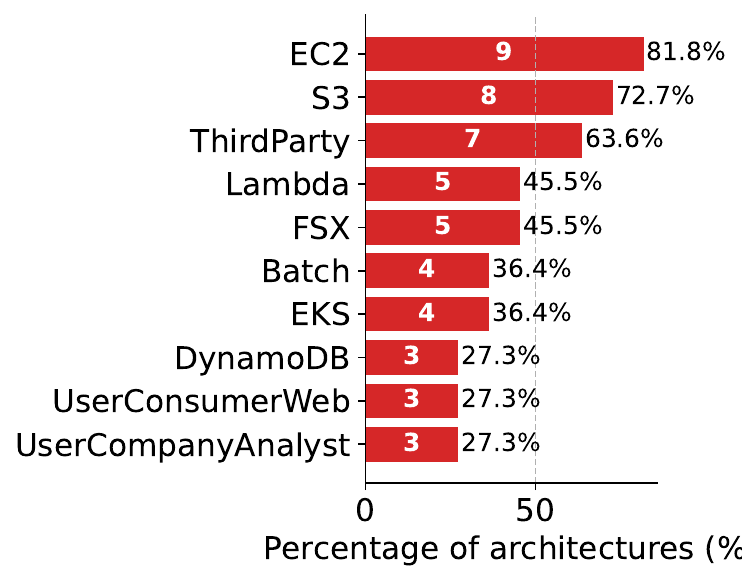}
    \caption{HPC}
    \label{fig:top10:HPC}
  \end{subfigure}
  \begin{subfigure}{0.25\textwidth}
    \includegraphics[width=\linewidth]{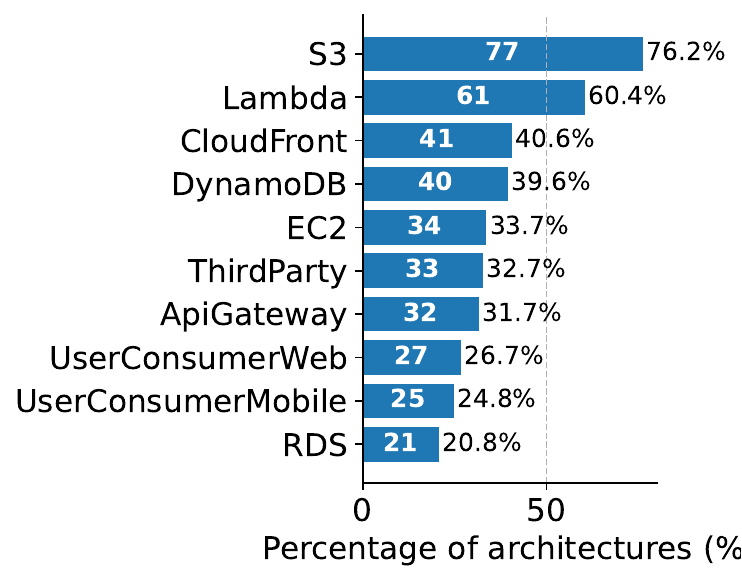}
    \caption{Edge}
    \label{fig:top10:Edge}
  \end{subfigure}%
  \begin{subfigure}{0.25\textwidth}
    \includegraphics[width=\linewidth]{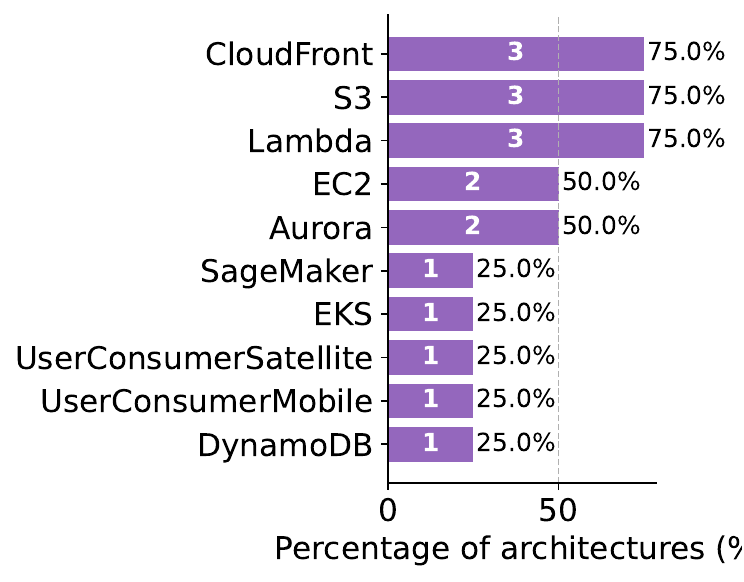}
    \caption{HPC+Edge}
    \label{fig:top10:HPC+Edge}
  \end{subfigure}%
  \begin{subfigure}{0.25\textwidth}
    \includegraphics[width=\linewidth]{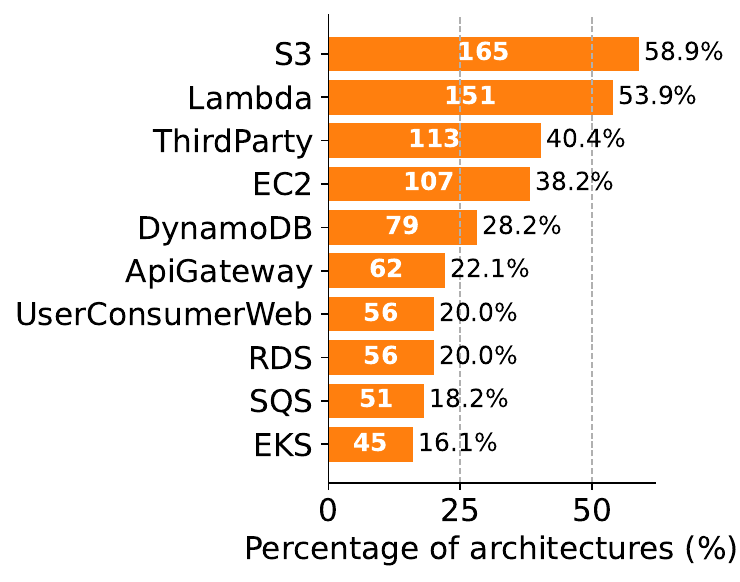}
    \caption{None}
    \label{fig:top10:None}
  \end{subfigure}
  \caption{Top ten most frequent services in HPC, edge and HPC+edge. The top 10 services in architectures outside those categories are provided for reference.}
  \label{fig:popular:services}
\end{figure*}

Fig.~\ref{fig:popular:services} shows the top ten most frequent services in each of the three types of architectures studied in this paper (see Table~\ref{tab:aws:services} for a description of each service).
We can observe that S3 is the most or second most popular service in all three types of architectures, appearing in more than 75\% of architectures for all three cases, though specific numbers differ.

\begin{table}[t]
    \centering
    \caption{Top AWS services in dataset plus other services in Fig.~\ref{fig:sample:architectures}-~\ref{fig:discriminating:services}. FS: File System; RDBMS: Relational database management system.}
    \label{tab:aws:services}
    \scriptsize
    \begin{tabular}{l|l}
    \hline
    \textbf{Service} & \textbf{Description} \\
    \hline
    S3 & Simple Storage Service, a serverless object storage service. \\
    Lambda &  Function-as-a-service (serverless compute). \\
    EC2 &  On demand virtual machines (can include images w/GPUs). \\
    DynamoDB & Serverless, NoSQL, fully managed database. \\
    API Gateway & Managed REST, HTTP, and WebSocket API gateway. \\
    CloudFront & Content delivery network (edge). \\
    FSx & High-performance FS:\,ONTAP, OpenZFS, Windows\,FS, Lustre. \\
    EFS &  NFS-compatible, serverless and elastic file storage. \\
    EKS &  Managed, elastic Kubernetes service. \\
    Batch &  Managed batch computing service for large-scale  workloads. \\
    Aurora &  RDBMS w/MySQL and PostgreSQL compatibility. \\
    StepFunctions & Visual workflow service for distributed applications. \\
    CloudWatch &  Service for monitoring and observability. \\
    SQS &  Serverless, distributed message queue. \\
    RDS &  Managed, relational database service. \\
    CloudFormation &  Infrastructure-as-Code (IaC), for managing cloud architectures. \\
    Service Catalog &  Curate and manage IaC cloud resources. \\
    IoT Core & Connect and communicate IoT devices w/ cloud applications. \\
    Lambda@Edge & Run cloud functions at the edge (CloudFront). \\
    EventBridge & Serverless asynchronous communication broker. \\
    OpenSearch & Managed OpenSearch (search and observability suite). \\ 
    Cognito & Customer identity and access management. \\
    Firehose & Ingest/transform/deliver data streams to data lakes and others. \\
    KDS & Kinesis Data Streams, collect and process realtime streams. \\
    AppSync &  Managed GraphQL API layer. \\
    SageMaker & Build, train and deploy machine learning models. \\
    \hline
    \end{tabular}
\end{table}

\begin{insightbox}
\refstepcounter{insightcounter}\textbf{Insight\,\theinsightcounter:}
S3 is the most or second most common service in all types of architectures, highlighting its versatility to serve workloads with diverse requirements.
\end{insightbox}

\begin{figure*}[t]
  \begin{subfigure}{0.31\textwidth}\centering
    \includegraphics[width=\linewidth]{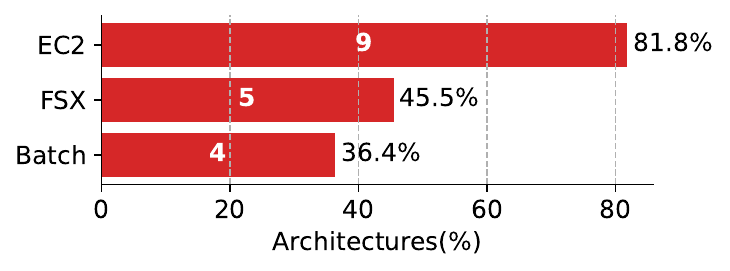}
    \caption{HPC}
    \label{fig:discriminating:HPC}
  \end{subfigure}%
  \begin{subfigure}{0.35\textwidth}
    \includegraphics[width=\linewidth]{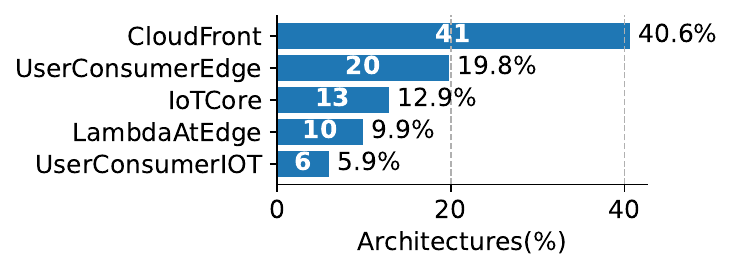}
    \caption{Edge}
    \label{fig:discriminating:Edge}
  \end{subfigure}%
  \begin{subfigure}{0.34\textwidth}
    \includegraphics[width=\linewidth]{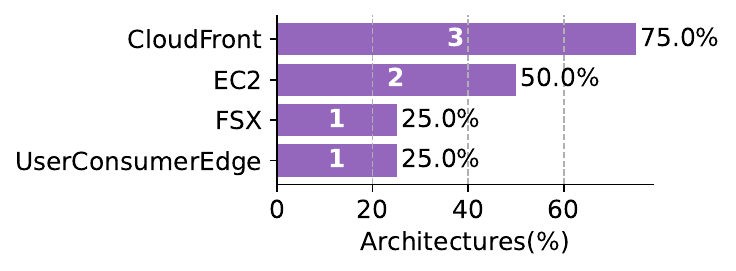}
    \caption{HPC+Edge}
    \label{fig:discriminating:HPC+Edge}
  \end{subfigure}
  \caption{Top HPC-specific and edge-specific services in each type of architecture.}
  \label{fig:discriminating:services}
\end{figure*}

In HPC architectures, EC2 is unsurprisingly the most common service (82\%), as computing clusters---built with virtual machines in the cloud---are essential to any HPC architecture~\cite{aws:hpc:lens}.
Next in popularity are FSx and Lambda (46\% each), Batch and EKS (36\% each) and DynamoDB (27\%).
Almost two thirds (64\%) use ThirdParty services while UserConsumerWeb and UserCompanyDataStream appear in 27\% of the architectures.
If we consider only the HPC-specific services, then the top services in these architectures are EC2 (82\%), FSx (46\%) and Batch (37\%), as shown in Fig.~\ref{fig:discriminating:services}.

\begin{insightbox}
\refstepcounter{insightcounter}\textbf{Insight\,\theinsightcounter:}\
In HPC, compute (EC2, Lambda, Batch) and high-performance storage\,(FSx) appear frequently, though the S3 general-purpose object storage service is more prevalent than the HPC-specific FSx.
\end{insightbox}

In edge architectures Lambda, appears in 60\% of the architectures;
other popular services are CloudFront (41\%), DynamoDB (40\%), EC2 (34\%), API Gateway (32\%) and RDS (21\%).
In addition, these architectures typically contain third-party and proprietary services and devices, which appear in the architectures as ThirdParty (33\%), UserConsumerWeb (27\%) and UserConsumerMobile (25\%), though ThirdParty services could also be directly related to edge computing. 
Out of these, the only one specific to the edge is CloudFront.
Other edge services like UserConsumerEdge, IoT Core, Lambda@Edge and UserConsumerIoT appear less frequently in the architectures in the dataset (20\%, 13\%, 10\% and 6\%, respectively; see Fig.~\ref{fig:discriminating:services}).
AWS's Storage Gateway and Greengrass appear in only 3\% and 2\% of the architectures, respectively.

\begin{insightbox}
\refstepcounter{insightcounter}\textbf{Insight\,\theinsightcounter:}\
Edge architectures contain services common in web/mobile architectures combined with edge components provided by AWS (CloudFront, IoT Core, Lambda@Edge) or in-house/third-party devices and services.
Other AWS edge-specific services like Storage Gateway and Greengrass appear much less frequently. 
\end{insightbox}

In contrast, non-HPC/non-edge architectures use storage and compute services that are common in web/mobile applications (S3, Lambda, EC2, DynamoDB, API Gateway, RDS, SQS and EKS) in combination with ThirdParty components (40\%). 

Regarding pairs of services that appear frequently together, we found that in edge architectures, (Lambda, S3), (CloudFront, S3) and	(DynamoDB, Lambda) appear in 44\%, 35\%, and 30\% of the edge architectures, respectively.
In HPC architectures, the co-occurrences that appear in more than one third of the applications are (EC2, S3): 55\%, (EC2, ThirdParty): 55\%, (S3, ThirdParty): 46\%, (EC2, FSx): 36\%, (Batch, ThirdParty): 36\%, and (EC2, Lambda): 36\%. 

\begin{figure}[t]
    \centering
    \includegraphics[width=0.9\linewidth]{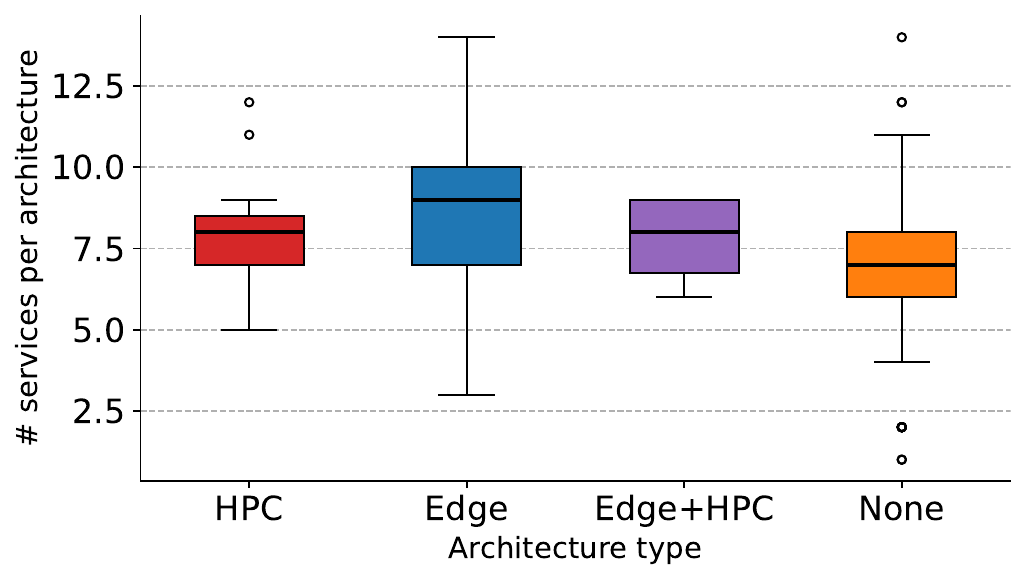}
    \caption{Number of services per architecture in each group.}
    \label{fig:uniqueAndAverage:services}
\end{figure}

We also assess how HPC and edge architectures differ with respect to the diversity of services used (Fig.~\ref{fig:uniqueAndAverage:services}).
On average, HPC architectures have slightly fewer services than edge (8.09 vs 8.50), while the architectures that belong to none of these groups have fewer services on average (7.04).

\begin{insightbox}
\refstepcounter{insightcounter}\textbf{Insight\,\theinsightcounter:}\
On average, HPC and edge architectures use 1-1.50 more services than other architectures (e.g., web), hinting at a higher design complexity.
\end{insightbox}

\subsection{Storage services (RQ3)}
\begin{figure*}[t]
  \begin{subfigure}{0.5\textwidth}\centering
    \includegraphics[width=8cm]{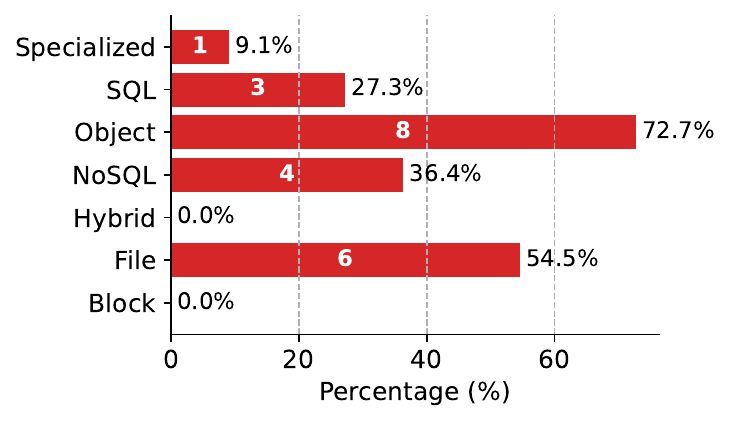}
    \caption{HPC}
  \end{subfigure}%
  \begin{subfigure}{0.5\textwidth}\centering
    \includegraphics[width=8cm]{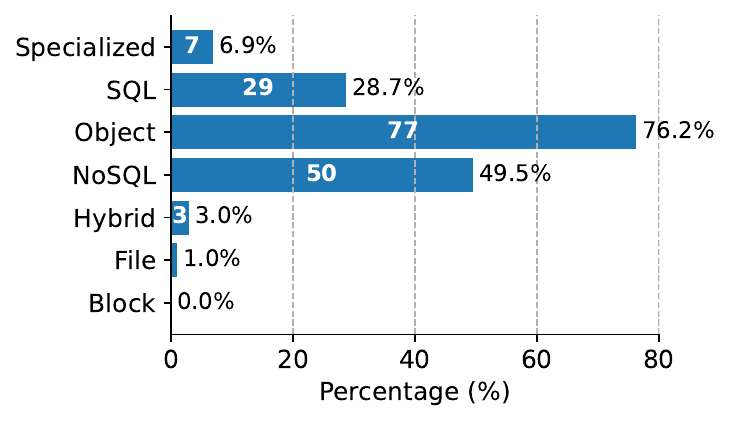}
    \caption{Edge}
  \end{subfigure}
  \begin{subfigure}{0.5\textwidth}\centering
    \includegraphics[width=8cm]{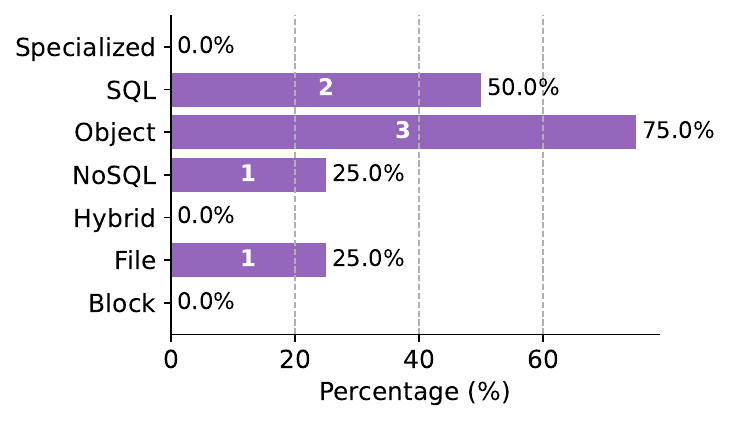}
    \caption{HPC+Edge}
  \end{subfigure}%
  \begin{subfigure}{0.5\textwidth}\centering
    \includegraphics[width=8cm]{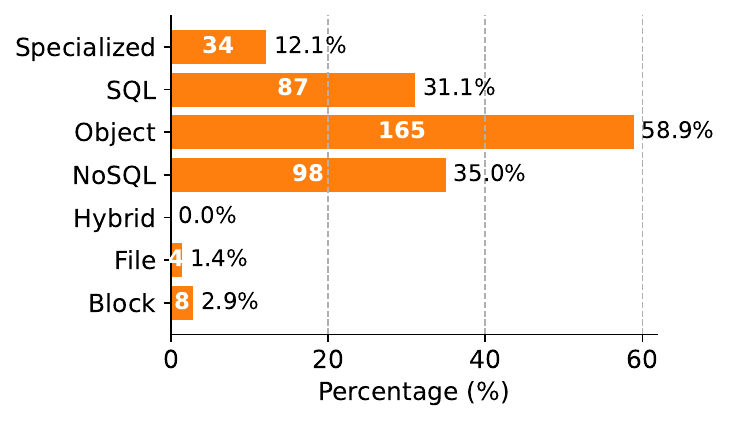}
    \caption{None}
  \end{subfigure}
  \caption{Types of storage services used in each type of architecture.}
  \label{fig:storage}
\end{figure*}

Storage services are a critical component of cloud architectures.
Within the context of this paper, we were specially interested in knowing what kind of storage services appear in HPC architectures; we extend the analysis to edge architectures for completeness.
Table~\ref{tab:aws:storage:services} describes the storage services analyzed in this subsection.
Fig.~\ref{fig:storage} shows the percentages of each storage service used in the studied groups.

We found that \emph{object storage} (S3) is the most used storage system, used in 73\%, 76\%, and 75\% of HPC, edge and HPC+edge architectures, respectively.
Surprisingly, S3 is only present in 60\% of the non-edge, non-HPC architectures;
we posit that this may be because in web/mobile applications database storage is much more relevant and thus S3 may not have been explicitly included in the architecture diagrams because the engineers describing the architecture did not see this component as an important one to include.
The Cloudscape authors~\cite{Satija:2025:Cloudscape} make a similar argument regarding EBS (Elastic Block Store): in practice this service is present in most architectures that use EC2 but not frequently seen in the dataset because developers rarely interface with it directly and, thus rarely mention it (2\% presence in the full dataset).

\emph{File systems} (FSx, EFS) are only popular in HPC architectures (55\%) and almost nonexistent in other scenarios (1\% edge, 1\% none), highlighting the importance of storing data in a high-performance solution like FSx for HPC.

Regarding \emph{databases}, SQL and NoSQL databases have about 1/3 prevalence in HPC (27\% and 36\%, respectively), whereas NoSQL is much more popular in edge solutions (50\% NoSQL, 29\% SQL).
Specialized databases (e.g., for analytics or time series data) are more popular in the non-edge, non-HPC architectures (12\%) than in HPC (9\%) or edge (7\%), possibly to handle Big Data use cases.

Finally, we note that AWS has a \emph{hybrid} edge+cloud storage solution, Storage Gateway, which is used in 3\% of edge architectures.
Not shown explicitly in the graphs, AWS's time series database, Timestream, appears in 1\% of the edge architectures but in none of the HPC ones, highlighting the importance of storing time series data (e.g., sensor readings) in some edge solutions.

\begin{table}[t]
    \centering
    \caption{Storage services; categories as defined in~\cite{Satija:2025:Cloudscape} and AWS's docs.}
    \label{tab:aws:storage:services}
    \begin{tabular}{l|l}
    \hline
    \textbf{Category} & \textbf{Services} \\
    \hline
    Specialized & RedShift, Neptune, Timestream \\
    SQL & RDS, Aurora \\
    Object & S3, MediaStore \\
    NoSQL & DynamoDB,\,DocumentDB,\,ElastiCache,\,MemoryDB \\
    Hybrid & Storage Gateway (hybrid edge+cloud) \\
    File system & EFS, FSx \\
    Block & EBS \\
    \hline
    \end{tabular}
\end{table}

\begin{insightbox}
\refstepcounter{insightcounter}\textbf{Insight\,\theinsightcounter:}\
File and object storage systems are critical components of HPC architectures, with many architectures combining multiple of these services to meet their requirements.
In contrast, edge architectures tend to combine object storage (S3) with one or more types of databases to meet their storage needs.
\end{insightbox}

\subsection{Machine learning services (RQ4)}
\begin{figure}[t]
    \centering
    \includegraphics[width=0.9\linewidth]{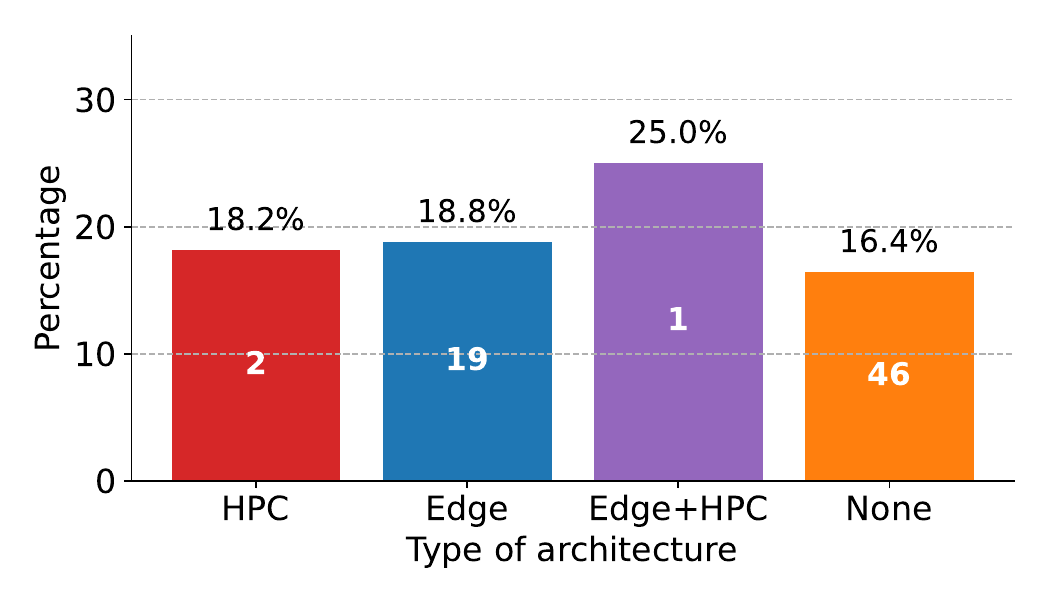}
    \vspace{-10pt}\caption{Percentage of architectures with 1+ ML services.}
    \label{fig:machine:learning:services}
\end{figure}

AWS has many machine learning (ML) services that go from general purpose ones like SageMaker, to very specific ones like Comprehend.
The list of ML services that appear at least once in the dataset is: SageMaker, SageMaker Ground Truth, Rekognition, Comprehend, Translate, Textract, Polly, Lex, Lookout for Vision, Transcribe, and Tendra.
We say that an architecture uses ML services if it uses any of these.
Our analysis shows that the percentage of architectures that use at least one ML service is as follows: HPC 18\%, edge 19\%, HPC+edge 25\%, none 16\% (see Fig.~\ref{fig:machine:learning:services}).

\begin{insightbox}
\refstepcounter{insightcounter}\textbf{Insight\,\theinsightcounter:}\
HPC and edge architectures use machine learning services with similar frequency, and slightly more frequently (2+ percentage points) than other architectures.
\end{insightbox}

\subsection{Functional goals (RQ5)}
Cloudscape includes information about the functional goals of each architecture according to the following categories: Compute intensive, data ingestion, interactive, control plane and other.
This information is not part of the original architectures but was manually assigned by the Cloudscape team, according to their interpretation of the information presented in each video. 
Table~\ref{tab:functional} shows what percentage of architectures have the functional goals mentioned before.
A column may add up to more than 100\% as some architectures have more than one goal (e.g., compute intensive and data ingestion).
As expected, the most popular functional goal of HPC architectures is compute intensive (50\%) and for edge, data ingestion (50\%).
Second in popularity is data ingestion for HPC (37\%) and interactive (44\%) for edge;
the latter, likely due to the frequent use of CDNs in interactive applications.

\begin{insightbox}
\refstepcounter{insightcounter}\textbf{Insight\,\theinsightcounter:}\
HPC architectures are built to support compute intensive workloads, with data ingestion also being an important element.
Edge architectures are built for data ingestion but also to support interactive applications.
\end{insightbox}

\begin{table}[t]
    \centering
    \caption{Functional goals of the architectures in each category.}
    \label{tab:functional}
    \scriptsize
    \begin{tabular}{l|r|r|r|r}
    \hline
    \textbf{Functional goal} & \textbf{HPC} & \textbf{Edge} & \textbf{HPC+Edge} & \textbf{None} \\
    \hline
    Compute intensive & \colorbox{Red}{50\%} & 15\% & \colorbox{Orchid}{75\%} & 12\% \\
    Data ingestion & 36\% & \colorbox{NavyBlue}{50\%} & 25\% & \colorbox{BurntOrange}{40\%} \\
    Interactive & 27\% & 44\% & 50\% & 29\% \\
    Control plane & 9\% & 8\% &  & 22\% \\
    Other &  & 4\% &  & 10\% \\
    \hline
    \end{tabular}
\end{table}

\subsection{Prevalence of HPC and edge architectures (RQ6)}
\begin{figure}[t]
    \includegraphics[width=\linewidth]{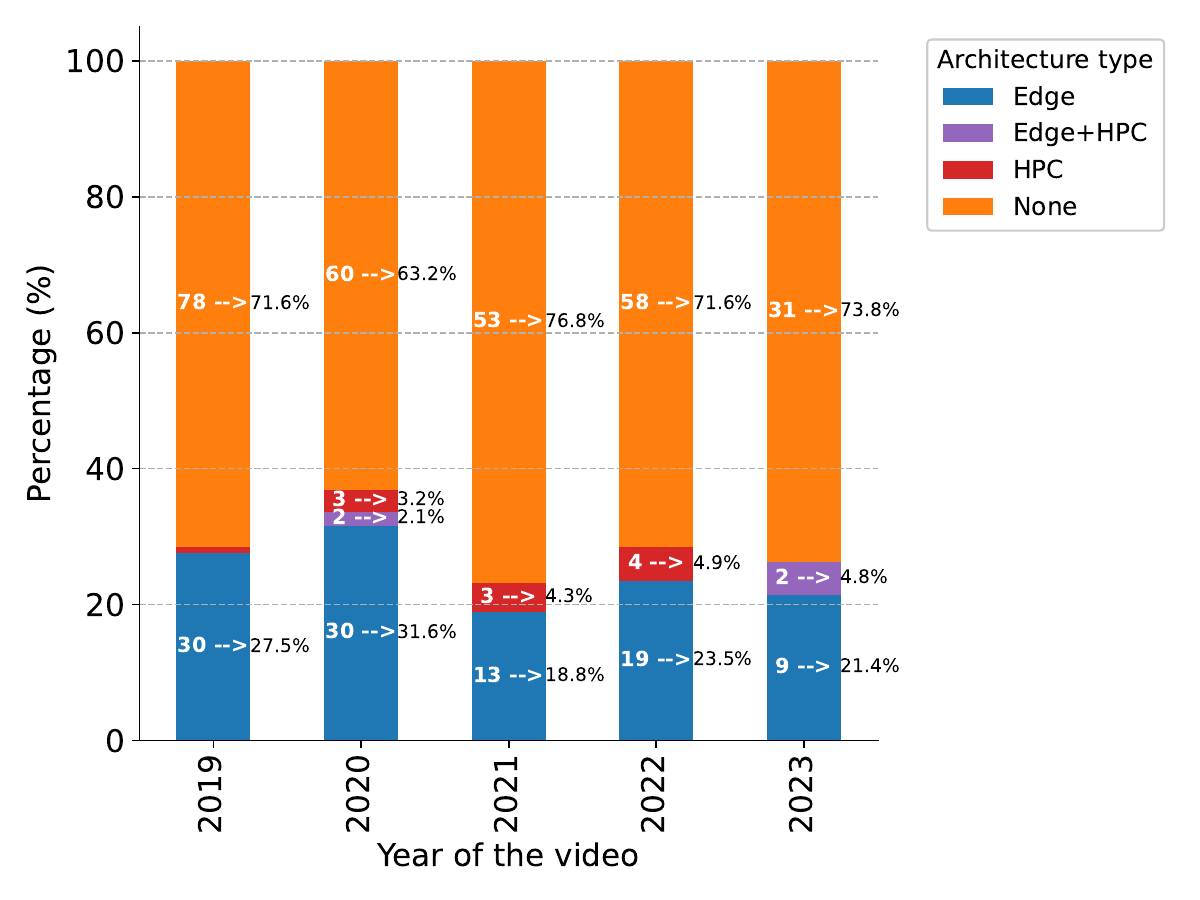}
  \vspace{-10pt}\caption{Per-year percentage of architectures in each category.}
  \label{fig:perYearPercentages}
\end{figure}

We find that the percentage of architectures in each of the categories varies every year (see Fig.~\ref{fig:perYearPercentages});
this distribution could occur by chance, or due to explicit human effort to include more (or less) architectures of a given type. 
For example, it is possible that the increase in HPC architectures observed in 2020--2022 is due to an explicit effort to document real scientific/parallel/HPC architectures in AWS.
Further, the purpose of the ``This is My Architecture'' series, the original source of the architectures in this study, is to showcase innovative architectures on AWS Cloud by customers and partners, so no claims of representativeness are made by its curators.

\subsection{Workflows (RQ7)}
\begin{figure}
    \centering
    \includegraphics[width=0.9\linewidth]{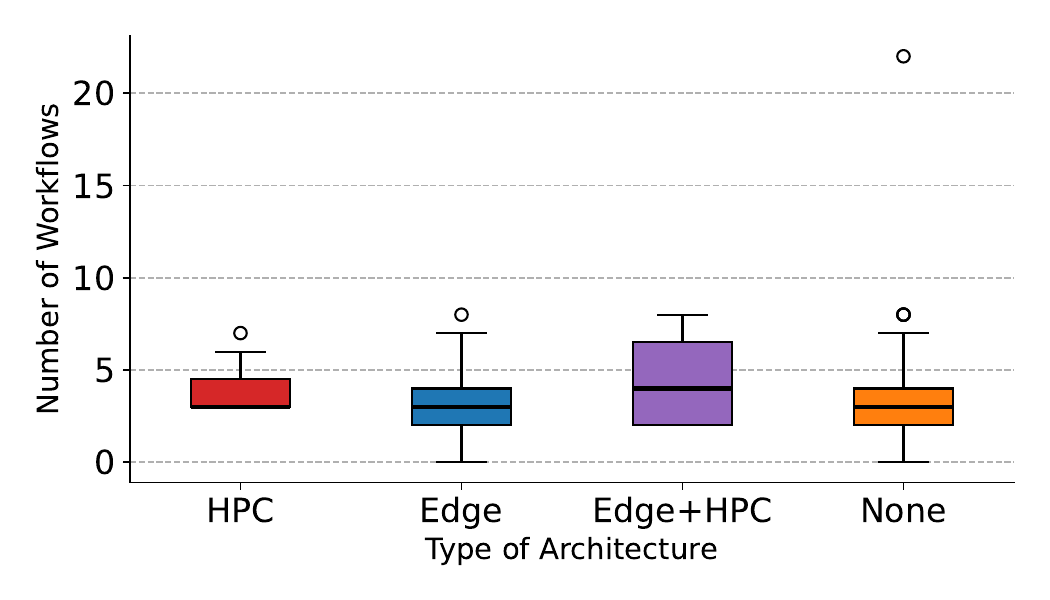}
    \vspace{-10pt}\caption{Boxplots of number of workflows per type of architecture. Mean: HPC=4.0, edge=3.3, HPC+edge=4.5, none=3.1.}
    \label{fig:workflows}
\end{figure}

The architectures in the dataset have edges that belong to one or more workflows.
These workflows are synchronous sequences of requests and data movements resulting from a trigger.
We find that non-edge, non-HPC architectures, have between 0 and 22 workflows, with an average of 3.12 (see Fig.~\ref{fig:workflows}).
In contrast, edge architectures have between 0 and 8 workflows, with an average of 3.27.
HPC architectures have between 3 and 7 workflows, with an average of 4.00.
These are architectural workflows, initiated by triggers and denoted by edges in the diagrams, and are not computational workflows which are not visible in an architecture diagram.

\begin{insightbox}
\refstepcounter{insightcounter}\textbf{Insight\,\theinsightcounter:}\
The difference in the number of workflows for the architectures in each group is not high, with all types having an average between 3.1 and 4.5.
\end{insightbox}

\begin{figure*}[t]
  \begin{subfigure}{0.4\linewidth}\centering
    \includegraphics[width=7.5cm]{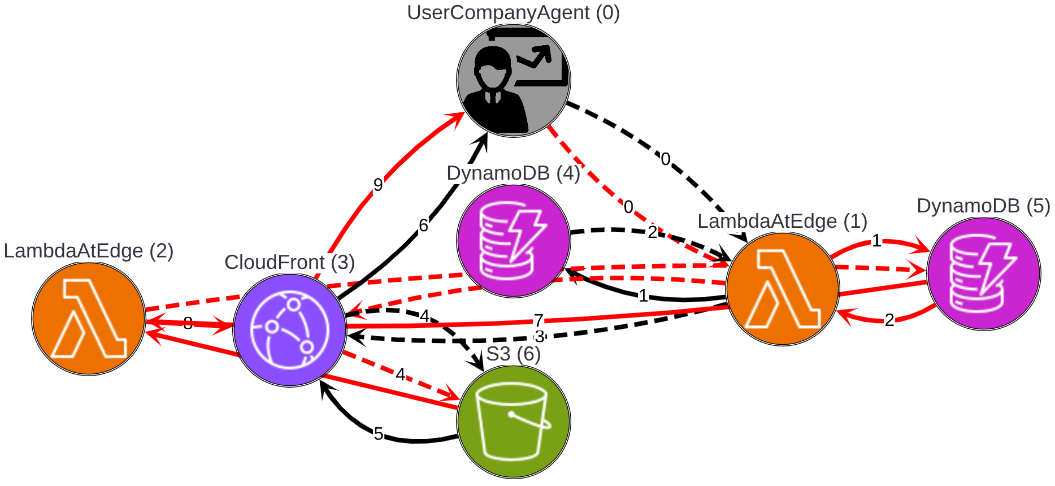}\vspace{15pt}
    \caption{Deep Instinct: RT Statistics \& Decisions w/\,Lambda@Edge.}
  \end{subfigure}\hspace{6pt}
  \begin{subfigure}{0.29\linewidth}%
  \centering
    \includegraphics[width=4.8cm]{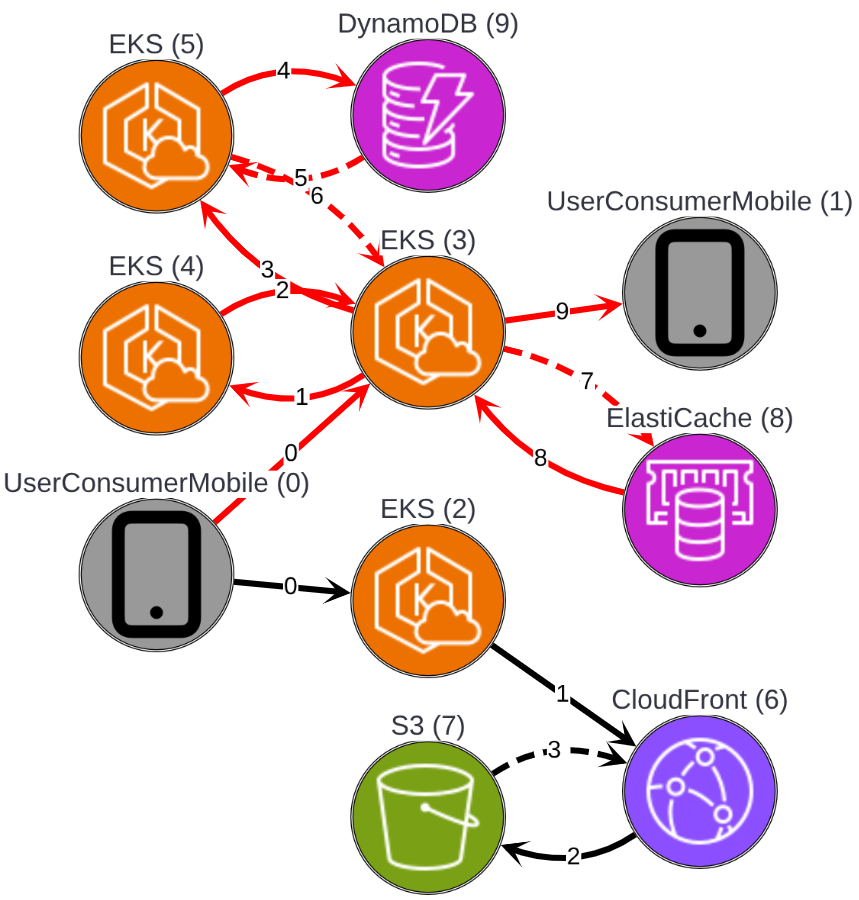}%
    \caption{Snap: Journey of a Snap on Snapchat.}
  \end{subfigure}\hspace{6pt}
  \begin{subfigure}{0.29\linewidth}
  \centering
    \includegraphics[width=4cm]{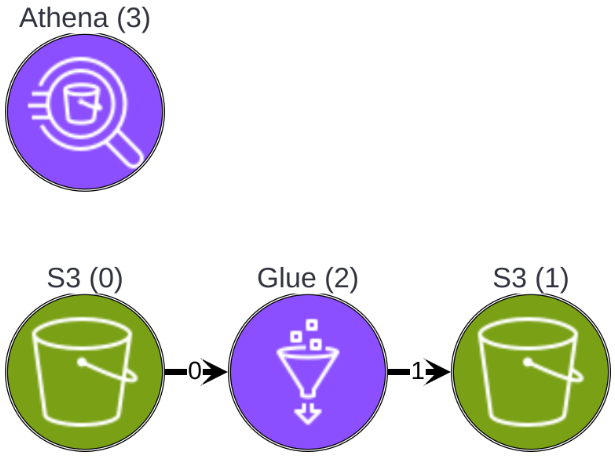}\vspace{20pt}
    \caption{iRobot: Serverless Data Cataloging.}
  \end{subfigure}
  \vspace{-10pt}\caption{Three representative edge applications.}
  \label{fig:edge:architectures}
\end{figure*}

\subsection{A deeper dive into edge architectures}
After answering RQ1-RQ2 about the services used in the architectures, we performed a more in-depth look into the edge group because it is the only one big enough to support a more detailed analysis and pattern finding.
We used $k$-means to try to find clusters with common architectural patterns.
We encoded each edge cloud architecture into a binary vector, with each dimension representing a specific AWS service.
To deal with the high dimensionality of these vectors and facilitate visualization, we applied Principal Component Analysis (PCA) to reduce the data to a 2D space.
This allowed us to visually inspect the data and any groupings identified by $k$-means.
We applied the clustering algorithm to the dimension-reduced data and used the elbow method to determine the optimal number of clusters ($k=4$), where the decrease in the sum of squared errors began to level off.
We then visually examined the results for both $k=3$ and $k=4$ and found no clearly identifiable clusters; a manual analysis of the top services in each cluster showed small differences in prevalence for each cluster but no interesting patterns emerged.
We are not including the results of this analysis in the paper but we have included it in our GitHub repository in case anyone finds it useful.

Finally, we used $k=1$ to find the most ``representative'' edge applications in the dataset.
We identified the 6 architectures closest to the cluster mean\footnote{IDs: tTQ36qQF\_vA, IV3KuMGVNXI, Cgv0kfp\_6xQ, GoziWpmFCS0, wjtSHyENv0I, Yju3yReAQtc. One can see the deails of each in YouTube using the ID, e.g.: \url{https://www.youtube.com/watch?v=IV3KuMGVNXI}} and visually inspected each.
Four of them contain the three top services in Fig.~\ref{fig:top10:Edge} (S3, Lambda, CloudFront), with CloudFront being the edge component; three of those also include the fourth more popular service in the group: DynamoDB.
A fifth one, the Snap architecture in Fig.~\ref{fig:edge:architectures}, contains three of the top four services (S3, CloudFront and DynamoDB).
We chose the two most interesting of those five ``representative'' architectures and show them in Fig.~\ref{fig:edge:architectures}.
The Deep Instinct architecture uses Lambda@Edge to make real-time cybersecurity decisions at the edge, like controlling rolling upgrades of client agents and modifying responses from the origin based on real-time data.
The Snap architecture contains an interesting upload path of photos from mobile devices to CloudFront via a media service managed by Kubernetes. 
Finally, the iRobot architecture is very different from the others, containing only the most popular service, S3.
This is one of the architectures that we manually tagged, as it does not contain any edge-specific service.
We included this architecture based on the description which explains how the IoT data is ingested into the data pipeline:  JSON real-time data from millions of robots is streamed into S3 and DynamoDB through a serverless ETL pipeline (SQS, Lambda, Kinesis, Kinesis Firehose) and landing in an S3 bucket (shown in the architecture).

\section{Threats to Validity}
\label{sec:threats:to:validity}
While we think that our curated dataset and analysis can be helpful in understanding how industry builds HPC and edge architectures in the cloud, there are limitations in the data and methodology that limit the generalizability of the results.

\vspace{6pt}
\paragraph*{\hspace{-20pt}Limitations from the original data}
The AWS team in charge of the ``This is My Architecture'' content may not seek to provide a representative sample of cloud architectures running on AWS.
Thus, their own goals and biases (e.g., as evidenced by documenting more HPC architectures in some years), and the willingness of industry participants to clearly and thoroughly describe their architectures, limits the generalizability of the dataset. 
For example, the limitations of the verbal descriptions of the architectures are evident in the minimal presence of EBS in the dataset; the Cloudscape team believes that the engineers in the videos thought mentioning EBS was not important or they forgot to do so, as they do not interact directly with it when storing data in EC2.
Our analysis was also limited by the fact that the main focus of the Cloudscape team was studying the use of storage services so more information was documented related to these than other services.

\vspace{6pt}
\paragraph*{\hspace{-20pt}Limitations in our curation process}
When classifying an architecture into one of the categories in this study, we did so by filtering by the names of the services the architecture contains, without any information on their configuration details; thus, we may have missed some architectures that fall into our groups of study because they use services that are not exclusive to edge or HPC.
Two notable cases are EC2, that can be part of a \emph{traditional cluster environment} (as described by AWS in their HPC Lens documentation~\cite{aws:hpc:lens}), and DataStream, which could represent IoT input datastreams but could also represent ClickStreams or (Big-)data streams. 
Further, our own biases may have affected the manual tagging based on the textual description of the architectures; however, we added very few architectures using this process, so any errors here have limited effects on the overall analysis.

\vspace{6pt}
\paragraph*{\hspace{-20pt}Limitations of our results}
While we found 100+ architectures that meet our edge inclusion criteria, HPC architectures constitute a small percentage of the Cloudscape dataset (11 HPC + 4 HPC+edge), limiting the generalizations that can be made from analyzing these two groups.
Nevertheless, even though with a larger sample the specific statistics could change, we believe that the general trends between the groups are likely to remain relatively stable, as our findings fit with our domain knowledge of how these architectures are built.

\section{Related Work}
\label{sec:related}
Our work builds upon the work by Satija et al.~\cite{Satija:2025:Cloudscape}, who compiled the original Cloudscape dataset.
We curated and extended this dataset and performed an analysis that looks at the data from a different angle.
Whereas the Cloudscape team wanted to learn more about the storage services in use, we look at the architectures that have HPC or edge components, and characterize them, answering seven research questions.

Similar to Cloudscape but sourced from GitHub data and some selected scientific use cases, Eismann et al.~\cite{Eismann:2021:Why,Eismann:2022:ServerlessCharacterization} analyzed real cloud applications, but only if they contained one or more serverless components.
Another dataset of serverless applications is Wonderless~\cite{Eskandani:2021:wonderless}, though the authors' focus was on collecting the dataset from GitHub and not on characterizing the architectures.
In another work from our group, we collected OpenLambdaVerse~\cite{Chavez:2025:OpenLambdaVerse}, an updated version of the Wonderless dataset and performed an extensive characterization of the repositories.
However, we did not look into how applications in the edge or HPC domains are being built.

WorkflowHub~\cite{Gustafsson:2025:WorkflowHub} is a recent registry of 800+ computational workflows.
Our dataset provides no insight into this type of workflow,
thus WorkflowHub is inherently different from our work, and both may be of interest to the HPC community.

Others have analyzed specific cloud architectures for edge~\cite{Moreno:2019:disaggregated,Li:2022:edge,Belcastro:2025:DroneMonitoring} and HPC~\cite{Mateescu:2011:HybridHPC,Castane:2018:Ontology,Imbrosciano:2025:HaMMon}, but these have only looked at single scenarios and not at a large set of real architectures.
Our study complements these works, with a novel characterization of real HPC and edge architectures in AWS.

\section{Conclusions}
\label{sec:conclusions}
We analyzed a recent dataset of 396 cloud architectures and generated a curated set of architectures that contain HPC or edge components.
To the best of our knowledge, this resulting dataset is the first dataset of real cloud architectures (in AWS) pertaining to these communities.
We characterized and analyzed the architectures, seeking to understand how industry builds their HPC and edge solutions in the cloud.
From this, we distil several insights that we hope will be useful to practitioners and researchers.
We have released our processing scripts to aid in reproducibility and facilitate re-use.

\balance
\bibliographystyle{IEEEtran}
\bibliography{bibliography}

\begin{thebibliography}{10}
\providecommand{\url}[1]{#1}
\csname url@samestyle\endcsname
\providecommand{\newblock}{\relax}
\providecommand{\bibinfo}[2]{#2}
\providecommand{\BIBentrySTDinterwordspacing}{\spaceskip=0pt\relax}
\providecommand{\BIBentryALTinterwordstretchfactor}{4}
\providecommand{\BIBentryALTinterwordspacing}{\spaceskip=\fontdimen2\font plus
\BIBentryALTinterwordstretchfactor\fontdimen3\font minus
  \fontdimen4\font\relax}
\providecommand{\BIBforeignlanguage}[2]{{%
\expandafter\ifx\csname l@#1\endcsname\relax
\typeout{** WARNING: IEEEtran.bst: No hyphenation pattern has been}%
\typeout{** loaded for the language `#1'. Using the pattern for}%
\typeout{** the default language instead.}%
\else
\language=\csname l@#1\endcsname
\fi
#2}}
\providecommand{\BIBdecl}{\relax}
\BIBdecl

\bibitem{aws:well:architected:framework}
{AWS Documentation}, ``{AWS} well-architected framework,''
  \url{https://docs.aws.amazon.com/wellarchitected/latest/framework/welcome.html},
  last accessed: June 20, 2025.

\bibitem{Satija:2025:Cloudscape}
S.~Satija, C.~Ye, R.~Kosgi, A.~Jain, R.~Kankaria, Y.~Chen, A.~Arpaci-Dusseau,
  R.~Arpaci-Dusseau, and Srinivasan, ``Cloudscape: {A} study of storage
  services in modern cloud architectures,'' in \emph{USENIX FAST}, 2025.

\bibitem{awsHPC}
{AWS Documentation}, ``High performance computing on {AWS},''
  \url{https://aws.amazon.com/hpc/}, last accessed: June 3rd, 2025.

\bibitem{awsEdge}
------, ``{AWS} for the edge,'' \url{https://aws.amazon.com/edge/services/},
  last accessed: June 3rd, 2025.

\bibitem{aws:hpc:lens}
------, ``High performance computing lens: Traditional cluster environment,''
  \url{https://docs.aws.amazon.com/wellarchitected/latest/high-performance-computing-lens/traditional-cluster-environment.html},
  last accessed: June 20, 2025.

\bibitem{Eismann:2021:Why}
S.~Eismann, J.~Scheuner, E.~van Eyk, M.~Schwinger, J.~Grohmann, N.~Herbst,
  C.~L. Abad, and A.~Iosup, ``Serverless applications: Why, when, and how?''
  \emph{IEEE Software}, vol.~38, no.~1, 2021.

\bibitem{Eismann:2022:ServerlessCharacterization}
S.~Eismann, J.~Scheuner, E.~v. Eyk, M.~Schwinger, J.~Grohmann, N.~Herbst, C.~L.
  Abad, and A.~Iosup, ``The state of serverless applications: Collection,
  characterization, and community consensus,'' \emph{IEEE Transactions on
  Software Engineering}, vol.~48, no.~10, 2022.

\bibitem{Eskandani:2021:wonderless}
N.~Eskandani and G.~Salvaneschi, ``The {W}onderless dataset for serverless
  computing,'' in \emph{IEEE/ACM Intl. Conf. Mining Softw. Repo. (MSR)}, 2021.

\bibitem{Chavez:2025:OpenLambdaVerse}
A.~Chavez-Moreno and C.~Abad, ``{OpenLambdaVerse}: {A} dataset and analysis of
  open-source serverless applications,'' in \emph{IEEE Intl. Conference on
  Cloud Engineering (IC2E)}, 2025.

\bibitem{Gustafsson:2025:WorkflowHub}
O.~Gustafsson, S.~Wilkinson, F.~Bacall, S.~Soiland-Reyes, S.~Leo, L.~Pireddu,
  S.~Owen, N.~Juty \emph{et~al.}, ``{WorkflowHub}: {A} registry for
  computational workflows,'' \emph{Scientific Data}, vol.~12, no.~1, 2025.

\bibitem{Moreno:2019:disaggregated}
R.~Moreno-Vozmediano, E.~Huedo, R.~S. Montero, and I.~M. Llorente, ``A
  disaggregated cloud architecture for edge computing,'' \emph{IEEE Internet
  Computing}, vol.~23, no.~3, 2019.

\bibitem{Li:2022:edge}
J.~Li, C.~Gu, Y.~Xiang, and F.~Li, ``Edge-cloud computing systems for smart
  grid: state-of-the-art, architecture, and applications,'' \emph{Journal of
  Modern Power Systems and Clean Energy}, vol.~10, no.~4, 2022.

\bibitem{Belcastro:2025:DroneMonitoring}
L.~Belcastro, C.~Cosentino, F.~Marozzo \emph{et~al.}, ``Empowering efficient
  drone monitoring with low-latency edge-cloud continuum platforms,'' in
  \emph{Euromicro Intl. Conf. Par., Distrib., and Network-Based Proc.}, 2025.

\bibitem{Mateescu:2011:HybridHPC}
G.~Mateescu, W.~Gentzsch, and C.~J. Ribbens, ``Hybrid computing—where {HPC}
  meets grid and cloud computing,'' \emph{Future Generation Computer Systems},
  vol.~27, no.~5, 2011.

\bibitem{Castane:2018:Ontology}
G.~G. Casta{\~n}{\'e}, H.~Xiong, D.~Dong, and J.~P. Morrison, ``An ontology for
  heterogeneous resources management interoperability and hpc in the cloud,''
  \emph{Future Generation Computer Systems}, vol.~88, 2018.

\bibitem{Imbrosciano:2025:HaMMon}
M.~Imbrosciano, E.~Sciacca, F.~Vitello, L.~Pelonero, F.~Franchina, U.~Becciani,
  I.~Colonnelli, and D.~Medic, ``{ The Cloud-HPC infrastructure for Hazard
  Mapping and vulnerability Monitoring (HaMMon) },'' in \emph{Euromicro Intl.
  Conf. Par., Distrib., and Network-Based Proc.}, 2025.

\end{thebibliography}

\end{document}